\documentclass[12pt]{article}
\usepackage{longtable}
\usepackage[dvips]{graphicx}
\begin{document}
\title{Gravitational quasinormal radiation of higher-dimensional black holes.}
\author{R.A. Konoplya \\
Department of Physics, Dniepropetrovsk National University\\
St. Naukova 13, Dniepropetrovsk  49050, Ukraine\\
$\mathrm{konoplya_{-}roma@yahoo.com}$}
\date{}
\maketitle
\thispagestyle{empty}

\begin{abstract}
We find the gravitational resonance (quasinormal) modes of the higher dimensional
Schwarzschild and Reissner-Nordstrem black holes. The effect on the quasinormal
behavior due to the presence of the $\lambda$ term is  investigated.
The QN spectrum is totally different for different signs of
$\lambda$. In more than four dimensions there excited three types
of gravitational modes: scalar, vector, and tensor. They produce three
different quasinormal spectra, thus the isospectrality  between
scalar and vector perturbations, which takes place for $D=4$
Schwarzschild and Schwarzschild-de-Sitter black holes, is broken in
higher dimensions. That is the scalar-type gravitational
perturbations, connected with deformations of the black hole horizon,
which damp most slowly and therefore dominate during late time of
the black hole ringing.
\end{abstract}


\section{Introduction and basic equations}

Two sorts of small perturbations of black holes within general relativity are
considered usually: the first consists in adding of a test field to a black hole
space-time and thus can be described by a dynamical equation for this field in
the background of a black hole. Second way to perturb a black hole is
to perturb the metric itself. In this case in order to find the evolution equation
one has to linearize the
Einstein equations. These perturbations are called
gravitational perturbations and are always paid more attention than
the first ones.
That is because gravitational radiation is much stronger than
that of the external fields decaying near the black hole, and also
the metric perturbations let us judge about stability of a black hole.

In four space-time dimensions the linearized Einstein equations
for perturbations of Schwarzschild black hole can be treated
separately for axial (invariant under the change
$\phi \rightarrow - \phi$) and polar perturbations.
After the separation of angular variables and supposing exponential dependence
of the wave function on time ($\Psi_{i} \sim e^{-i \omega t}$), the final radial
wave equations for these two kinds of perturbations have the form of the
Schrodinger wave-like equations with some different effective potentials
\begin{equation}\label{1}
\left(\frac{d^{2}}{dr_{\ast}^{2}} + \omega^{2}\right) \Psi_{i}(r) = V_{i}
\Psi_{i}(r),
\end{equation}
where the index $i$ designates a type of perturbations.

It is a remarkable point that the spectra produced by
these two potentials for $D=4$ Schwarzschild black hole are the same.
What occurs for higher dimensional black hole was the challenge up
until now, when Ishibashi and Kodama \cite{Ishibashi1}, \cite{Ishibashi2},
\cite{Ishibashi3}
managed to reduce the
perturbation equations to three wave-like equations, one for each
type of perturbations: scalar (reducing to polar at $D=4$), vector
(reducing to axial at $D=4$), and "new" tensor perturbations.
Such a reduction of the linearized Einstein equations to
the wave-like form is not trivial work and requires decomposition of
the wave functions into the corresponding harmonics which are
connected with the symmetry of the background to be perturbed (see for example
references  in \cite{old-konoplya}).

The metric of the spherically symmetric charged static black hole in a
D-dimensional space-time with $\lambda$ term has the form:
\begin{equation}\label{2}
ds^{2}= -f(r) dt^{2} + f^{-1}(r)dr^{2} +r^{2}d\Omega^{2}_{D-2},
\end{equation}
where
\begin{equation}\label{3}
f(r)= 1-\frac{2 M}{r^{D-3}}+\frac{Q^{2}}{r^{2 D-6}}-\lambda r^{2}
\end{equation}
The effective potentials for scalar, vector, and tensor, gravitational
perturbations respectively have the form:
\begin{equation}\label{4}
V_{s}(r)=\frac{f (r) U(r)}{16 r^{2} \left(m + \frac{d (d+1)x}{2} -
\frac{d^2 Q^2}{r^{2 d -2}}\right)^{2}} ,
\end{equation}
where
$$U(r)=[-d^{3} (d+2) (d+1)^{2} x^{2}+4 d^{2} (d+1) (d (d^2 + 6 d -4) z+ 3 (d-2)m
)x- $$
$$12 d^5 (3 d -2) z^2 - 8 d^2 (11 d^2 -26 d +12) m z + 4 (d-2) (d-4) m^{2}]\lambda
r^{2}+$$
$$d^{4} (d+1)^{2} x^{3}+d (d+1)[-3 d^2 (5 d^2 -5 d +2) z +4 (2 d^{2} -3 d +4) m + $$
$$d (d-2)(d-4)(d+1)] x^{2} + 4 d [d^2 (4 d^3 +5 d^2 -10 d +4 ) z^2 - $$
$$ (d (34 - 43 d +21 d^2) m + d^2 (d+1) (d^2 -10 d + 12) ) z - 3 (d - 4) m^2 - $$
$$ d (d+1)(d-2) m ] x - 4 d^5  (3 d - 2) z^3 + 12 d^2 (2 (-6 d + 3 d^2 +4)
m+$$
$$ d^2 (3 d - 4) (d-2)) z^2 + (4 (13 d -4) (d-2) m^2 + 8 d^2 (11 d^2 -18 d + 4) m) z+$$
\begin{equation}\label{5}
16 m^{3} +4 d (d+2) m^{2},
\end{equation}
\begin{equation}\label{6}
x=\frac{2 M}{r^{d-1}}, \qquad d=D-2, \qquad m=l(l+d-1) -d, \qquad z =
\frac{Q^2}{r^{2 d -1}}.
\end{equation}
\begin{equation}\label{7}
V_{v\pm}(r)=\frac{f(r)}{r^{2}}\left( l (l+d-1) - 1+\frac{d^2 - 2 d +
4}{4}- \frac{d (d-2) \lambda r^2 }{4} + \frac{d (5 d-2) Q^2}{4 r^{2 d -2}}
+\frac{\mu_{\pm}}{r^{d-1}}\right),
\end{equation}
\begin{equation}\label{7a}
\mu_{\pm} = - \frac{d^2 +2}{2} M \pm \sqrt{((d^2 -1)^{2} M^{2} + 2 d (d-1)
(l (l+d-1) - d) Q^{2})}
\end{equation}
\begin{equation}\label{8}
V_{t}(r)=f(r)\left(\frac{l(l+d-1)}{r^2} + \frac{(d) (d-2)}{4 r^2} f(r) +
\frac{d}{2 r} f'(r)\right),
\end{equation}

The effective potential for tensor perturbations proved to be equivalent
to that corresponding to decay of test scalar field in a black hole
background \cite{Konoplya-prd3},\cite{Hartnoll}.
When $\lambda > 0$ $(< 0)$  one has D-dimensional SdS (SAdS) background.
Note that the scalar-type perturbation equations for charged background
together with the corresponding Maxwell equations can be reduced to
the pair of equations for electromagnetic and gravitational
perturbations (see eqs. (5.59) - (5.63) of \cite{Ishibashi3}), which
we shall not write out here, since they are cumbersome.

When perturbing a black hole,  its response
has resonances at some complex discrete modes.
If one denote the general solution at infinity as
\begin{equation}\label{9a}
\Psi = A_{in} \psi_{in} + A_{out} \psi_{out}, \qquad r^{*} \rightarrow \infty
\end{equation}
where $r^{*}$ is the tortoise coordinate ($d r^{*} = d r/f(r)$), $\psi_{in} \sim e^{-i \omega r^{*}}$
$\psi_{out} \sim  e^{i \omega r^{*}}$  are radial solutions
corresponding to incoming and outgoing radiation at infinity with
some complex amplitudes $A_{in}$,  $A_{out}$.
Then the reflection coefficient $A_{out}/A_{in}$ is singular at the resonance frequencies.
Therefore one can think that $A_{in}$ vanishes under the non-vanishing $A_{out}$, i.e. no
incoming radiation is permitted at spacial infinity.
Since no outgoing radiation can escape the horizon
of a black hole, the modes are required to represent purely
in-going waves at the horizon. Thus under the choice of positive
sign of the real part of $\omega$, $\omega= \omega_{Re}-i
\omega_{Im}$, ($\Psi_{i} \sim e^{-i \omega t}$)
the resonance (quasinormal) modes satisfy the boundary conditions:
\begin{equation}\label{4}
\Psi \sim c_{\pm} e^{\pm i \omega  r^{*}} \qquad as \qquad  r^{*}\rightarrow \pm
\infty.
\end{equation}
It is important that the quasinormal modes dominate during late-time
stages of the gravitational response of a black hole to an external
perturbation. This response consists of damping oscillations
and is called quasinormal ringing. Note that the quasinormal modes
do not depend on way in which they where excited and are determined
by a black hole parameters only. Being calculated within the linear
approximation of metric perturbations, the existence of
quasinormal modes are confirmed in a non-linear analysis. They are
important characteristic values of various dynamic processes, such
as black hole collisions or decay of different fields in a BH
background.

In addition  it has been revealed recently that the
quasinormal modes calculated in AdS gravity have a direct
interpretation in the dual gauge theory \cite{horowitz} and, also, help to
predict the correct value of a black hole entropy in Loop quantum
gravity \cite{hoddreyer}. Al this stimulated considerable interest
in QNMs (see for example \cite{qn} and references therein). At the same time
the higher dimensional brane models \cite{dimopolos},
created new motivations to calculate quasinormal modes of the
higher-dimensional black holes \cite{CDL}, \cite{Konoplya-prd3},
\cite{Ida}, \cite{Berti}.

Thus within different contexts the quasinormal modes corresponding to
decay of a free scalar field around the higher dimensional ($D>4$) black holes have
been considered recently in \cite{Molina}, \cite{Motl-Neitzke}, \cite{horowitz},
\cite{konoplya}, \cite{CDL}, \cite{Konoplya-prd3}, \cite{Ida}.
We are interested here in calculating of the quasinormal modes of the
D-dimensional S, SdS and SAdS black holes for gravitational perturbations.
In Sec.II we calculate the QNMs of $D>4$ SBH. In Sec.III the case
of non-vanishing $\Lambda$ - term is investigating; this includes
black holes in de-Sitter and Anti-de-Sitter space-times.
Sec. IV is devoted to the effect of the
black hole charge on the QN ringing.

\section{Schwarzschild}

The effective potentials for vector and tensor potentials for Schwarzschild
black holes (see formulas (\ref{7}) and (\ref{8}) at $Q=0$)
is constant at the infinity and at the horizon, and approach
a maximum somewhere in between (see \cite{Ishibashi2} for plots of
the effective potentials). Together with the boundary conditions (\ref{4}) this
allows us to use the standard WKB approach of Schutz and Will \cite{Will-Schutz}
which have been recently extended to the 6th order \cite{Konoplya-prd3}.
Another case is the scalar-gravitational potential (\ref{4}) which has negative
pitch near the event horizon for some values of the parameters $D$ and $l$.
Thus when analyzing this particular type of gravitational perturbations we
shall be restricted by those values of $D$ and $l$ for which we have
good potential barrier for $V_{s}$.

In \cite{Konoplya-prd3} it was shown that the real part of quasinormal modes
corresponding to decay of the free scalar field around the D-dimensional
Schwarzschild black hole are roughly proportional to $D$ if
measured in units $2 r_{0}^{-1}$ where $r_{0}$ is the horizon radius,
and for moderate and large $D$ this numerical linear dependence is
very accurate.

These QNMs of free scalar field coincide completely
with those of the tesor-type gravitational perturbations,
which is naturally, since the effective potentials of these two
perturbations are equivalent.
To compute the QNMs we used the 6th order WKB formula
\cite{Konoplya-prd3}:
\begin{equation}\label{100}
\frac{\imath Q_{0}}{\sqrt{2 Q_{0}''}}
- \sum_{i=2}^{i=6} \Lambda_{i} =n+\frac{1}{2},
\end{equation}
where the correction terms of the i-th WKB order $ \Lambda_{i}$
can be found in \cite{Will-Schutz} and \cite{Konoplya-prd3}, $Q = V - \omega^2$
and $Q_{0}^{i}$ means the i-th derivative of $Q$ at its maximum.
The fundamental modes for vector and tensor $l=2$ potentials are presented in Table 1
(the scalar potential for $l=2$ has negative pitch in higher dimensions).

\begin{table}
\begin{center}
\begin{tabular}{|c|c|c|}
\hline
  D & tensor-type & vector-type \\ \hline
  4 & - & 0.3736 - 0.0890 i \\
  5 & 1.0681 - 0.2529 i & 0.8056 - 0.2355 i \\
  6 & 1.5965 - 0.3987 i & 1.2249 - 0.3781 i \\
  7 & 2.0998 - 0.5317 i & 1.6542 - 0.5110 i \\
  8 & 2.5910 - 0.6536 i & 2.0865 - 0.6283 i \\
  9 & 3.0760 - 0.7663 i   & 2.5231 - 0.7364 i \\
  10 & 3.5574 - 0.8717 i & 2.9539 - 0.8369 i \\
  11 & 4.0354 - 0.9721 i & 3.4085 - 0.9311 i \\
  12 & 4.5092 - 1.0697 i & 3.8563 - 1.0197 i \\
  13 & 4.9767 - 1.1668 i & 4.3069 - 1.1036 i \\
  14 & 5.4348 - 1.2661 i & 4.7601 - 1.1833 i \\
  15 & 5.8797 - 1.3700 i & 5.2154 - 1.2594 i \\ \hline
\end{tabular}
\end{center}
Table I. The fundamental QN frequencies for $l=2$ gravitational perturbations
of SBH. The tensor-type gravitational mode for $D=4$ black hole does not exist.
\end{table}

\begin{table}
\begin{center}
\begin{tabular}{|c|c|c|c|}
\hline
  D & tensor-type & vector-type & scalar-type \\ \hline
  4 & - & 0.5994 - 0.0927 i & 0.5994 - 0.0927 i\\
  5 & 1.4198 - 0.2516 i & 1.2197 - 0.2362 i & 1.1345 - 0.2211 i\\
  6 & 2.0471 - 0.3959 i & 1.7356 - 0.3742 i & 1.5085 - 0.3029 i\\
  7 & 2.6192 - 0.5272 i & 2.2218 - 0.5055 i & 1.8767 - 0.4548 i\\
  8 & 3.1621 - 0.6467 i & 2.6993 - 0.6262 i & 2.2887 - 0.5517 i\\
  9 & 3.6891 - 0.7555 i & 3.1723 - 0.7368 i & 2.6781 - 0.6435 i\\
  10 & 4.2082 - 0.8541 i & 3.6426 - 0.8391 i& - \\
  11 & 4.7251 - 0.9427 i & 4.1116 - 0.9345 i& - \\
  12 & 5.2440 - 1.0216 i & 4.5800 - 1.0241 i& - \\
  13 & 5.7685 - 1.0907 i & 5.0483 - 1.1087 i& - \\
  14 & 6.3017 - 1.1504 i & 5.5168 - 1.1890 i& - \\
  15 & 6.8460 - 1.2009 i & 5.9857 - 1.2655 i& - \\ \hline
\end{tabular}
\end{center}
Table II. The fundamental QN frequencies for $l=3$ gravitational perturbations
of SBH. The effective potential of scalar-type modes has negative pitch when $D>9$ at
$l=3$, and therefore the WKB approach is inapplicable.
\end{table}

For $l=3$ we can obtain with the help of the WKB method the QNMs for
all three types of gravitational perturbations. Yet scalar-type effective potential
has negative pitch at $D>9$, and, what is more for WKB treatment, there is
another maximum near the event horizon, so that one would have to consider
the secondary scattering process near the peak of this sub-minimum.
Nevertheless for sufficiently large $D$ we can anticipate
the quasinormal behavior, since the horizon radius
$r_{0}$ does not depend on $D$ in this regime, $f(r)\rightarrow 1$
and one can re-scale
the wave equation, so that both real and imaginary QNMs are proportional to
$D$ \cite{Zaslavskii2}.

We see that the the more the spin weight of the type of
gravitational perturbations to be considered, the more the oscillation
frequency and the more the damping rate. Thus that is the scalar
type of perturbations that will dominate during  the later stages of
the quasinormal ringing. That seems to be connected with the fact
that, the scalar-type gravitational perturbations
describe the brane deformation
while the other two are just  a worldsheet
diffeomorphism
if one considers a vacuum brane.
It can be explicitly shown that the fluctuation of a vacuum
brane is completely described by the master variable for the scalar
mode of gravitational perturbations \cite{Ishibashi4}.

The real part of $\omega$ behaves similar to that of the
perturbations of the free scalar field \cite{Konoplya-prd3} showing
for moderate and large $D$ linear dependence on $D$ with good accuracy.

\section{$\Lambda$-term}

\subsection{Schwarzschild-de-Sitter}

The effective potentials for Schwarzschild black holes in higher dimensional
de-Sitter space-times behave similar to those for
asymptotically flat black holes: they approach the constants both
at infinity and at the horizon, and have maximum somewhere outside
the horizon. The only exception is the scalar-type gravitational perturbations
with some values of $D$ and $l$. Thus taking into account these exceptional
cases, we are able to apply the WKB formula of the previous section again.

The presence of positive cosmological constant changes the spectrum
of resonance oscillations in the following way: the real part of the
oscillations and the damping rate are decreasing as the $\Lambda$
is growing from $0$ to its extremal value $3 (D-3)/(D-1)^{\frac{D-1}{D-3}} M^{2}$.
We take here
$M=1$. The typical picture of dependence of the real and imaginary parts of
$\omega$ as a function of $\Lambda$ is shown on Fig.1 and Fig.2.

The QN modes of the Schwarzschild-de-Sitter black hole have been
considered in a number of papers (see for example \cite{mossnorman},
\cite{Molina}, \cite{cardosolemos1}, \cite{zhidenko1} and reference therein).
In the regime of th near extremal value of $\Lambda$,
which corresponds to the maximal mass of the black hole which can be
embedded into de-Sitter space-time
\begin{equation}\label{5}
M_{max}=\sqrt{\frac{(3/ \Lambda)^{D-3} (D-3)^{D-3}}{(D-1)^{D-1}}},
\end{equation}
the effective potential approaches the P\"{o}schl-Teller potential and the
quasinormal modes are best described by the P\"{o}schl-Teller analytic formula
\cite{cardosolemos1}. It have been shown also that the
sixth order WKB formula gives the values which are in excellent agreement
with those of P\"{o}schl-Teller in the near extremal limit \cite{zhidenko1}.

In addition, similar to SBH behavior, the more the spin weight of the type of
gravitational perturbations to be considered, the more the oscillation
frequency and the more the damping rate.

\begin{figure}
\begin{center}
\includegraphics{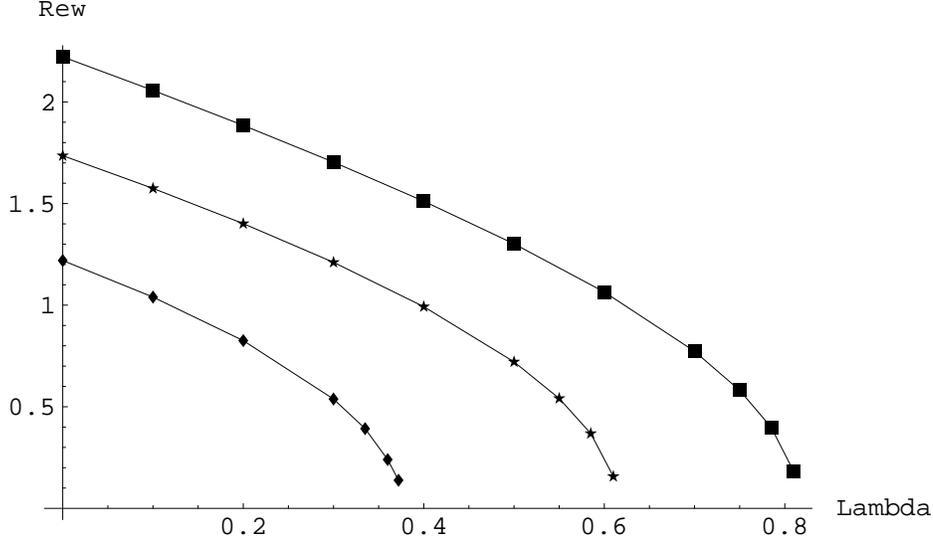}
\caption{SdSBH: $\omega_{Re}$ as a function of $\Lambda$ for vector-type
modes in the 6th order WKB approximation, $l=3$ $n=0$; $D=5$ (bottom),  $6$, $7$ (top).}
\label{}
\end{center}
\end{figure}

\begin{figure}
\begin{center}
\includegraphics{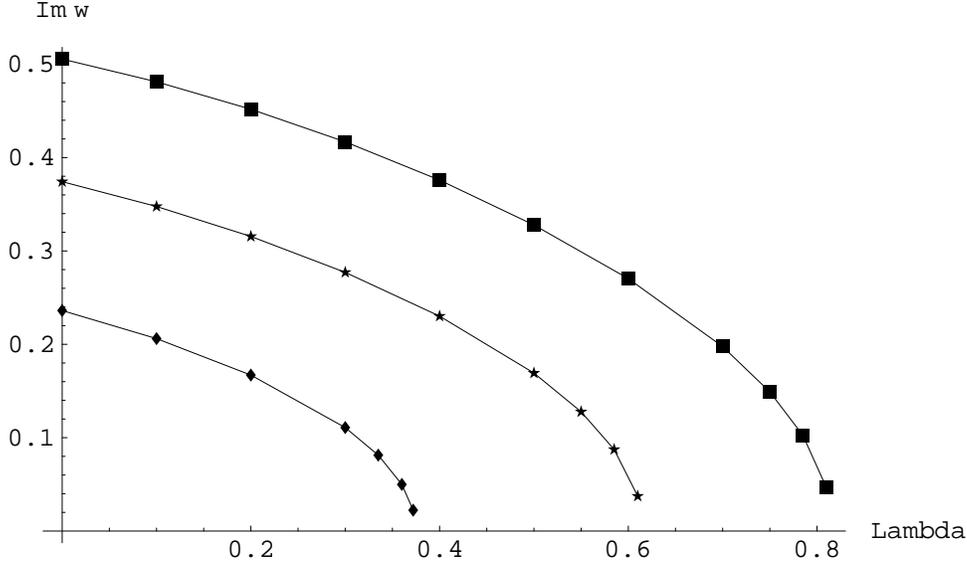}
\caption{SdSBH: $\omega_{Im}$ as a function of $\Lambda$ for vector-type
modes in the 6th order WKB approximation, $l=3$ $n=0$; $D=5$ (bottom),  $6$, $7$ (top).}
\label{}
\end{center}
\end{figure}

\subsection{Schwarzschild-anti-de-Sitter}

In AdS space-time there are two parameters which determine the
Schwarzschild black hole properties: the horizon radius $r_{+}$, and
the AdS radius $R= \sqrt{3/ \Lambda}$.
As is known from an extensive previous study \cite{horowitz}, \cite{qn}, the quasinormal
behavior of a black hole in anti-de-Sitter space-time crucially
depends upon the size of the black hole relative to the AdS radius
$R$. In particular, for large black holes ($r_{+} \gg R$) the QNMs are
proportional to $r_{+}$, and, thereby, to the black hole temperature
$T$ \cite{horowitz}, while for small black holes ($r_{+} \ll R$) they approach their
pure anti-de-Sitter values as $r_{+} \rightarrow 0$ \cite{konoplya}.
Asymptotically, for large overtone number $n$,
the QNMs do not depend on the spin of the field being considered
and are evenly spaced in $n$, no matter the size of the black hole \cite{CKL}.

By rescaling of $r$ we can put $R=1$. The effective
potentials (\ref{4}), (\ref{7}), (\ref{8}) are infinite at spatial
infinity. This lead to natural boundary condition: $\Psi=0$ at
infinity. In the context of $ADS_{d+1}/CFT_{d}$ correspondence,
this boundary condition, being suitable for free massless scalar field,
does not guarantee the coincidence with the poles of the
corresponding correlation function in the dual gauge theory for
gravitational perturbations
\cite{grav-boundary}.
Thus  which boundary conditions one should impose on gravitational
perturbations for the purpose of the CFT is an open question now, \cite{grav-boundary},
\cite{mossnorman}.

Within accepted Dirichle boundary condition, the wave function
vanishes at infinity and satisfies the purely in-going
wave condition at the black hole horizon.
Then one can compute the quasinormal frequencies stipulated by the
above potentials following the procedure of G.Horowitz and V.Hubeny
\cite{horowitz}.
The main point of that approach is to
expand the solution to the wave equation $\Psi $ around
$x_{+}=\frac{1}{r_{+}}$ ($x=1/r$):
\begin{equation}\label{18}
\psi (x)=\sum_{n=0}^{\infty} a_{n}(\omega)(x-x_{+})^{n}
\end{equation}
and to find the roots of the equation $\Psi(x=0)=0$ following from
the boundary condition at infinity. In fact, one has to truncate the
sum (\ref{18}) at some large $n=N$ and check that for greater values of $n$ the
roots converge.

Since one can anticipate the QN behavior for small black holes, and
for higher overtones \cite{CKL}, we shall consider here only
lowest overtones of large black holes. The data for fundamental QNMs  $D=5$
BH are collected in Table 1. The more $D$ the more time consuming is
computing of the quasinormal modes if requiring good accuracy,
since one has truncate the sum (\ref{18}) at larger $N$. Estimations
show that for higher $D$ the $n=0$ modes of scalar, tensor
types of gravitational perturbations and $n=1$ of the vector-type
perturbations are roughly the same for large black holes.
Since the purely decaying vector-type $n=0$ mode do not contribute into the
black hole ringing, i.e. do not oscillating, we find
that, similar to SBH and SdS BH ringing behavior, the scalar-type
oscillating modes have the least damping rate and thereby dominate
at late times of the ringing.

For $D=4$ SAdS black hole it was found an analytic expression for the
fundamental modes of vector-type (odd) gravitational perturbations \cite{CKL}:
\begin{equation}\label{19}
\omega_{n=0}= \frac{(l-1) (l+2)}{3 r_{+}} i.
\end{equation}
From our numerical data, for arbitrary dimensional AdS black hole
this formula can be generalized as follows
\begin{equation}\label{20}
\omega_{n=0}= \frac{(l-1) (l+D-2)}{(D-1) r_{+}} i.
\end{equation}

\begin{table}
\begin{center}
\begin{tabular}{|c|c|c|c|c|}
\hline
  $r_{+}$    &  scalar ($n=0$)   & vector ($n=0$) &  vector ($n=1$) & tensor ($n=0$)\\
\hline
  100     &  312.028 - 274.581 i   & 0 - 0.012489 i & 311.969 - 274.663 i & 311.982 - 274.659 i   \\
\hline
  150      & 467.973 - 411.944 i  & 0 - 0.008329 i & 467.933 - 411.999 i & 467.942 - 411.996 i   \\
\hline
 200      & 623.930 - 549.292 i    & 0 - 0.006249 i & 623.902 - 549.333 i & 623.909 - 549.331 i  \\
\hline
 250      & 779.896 - 686.634 i    & 0 - 0.00500 i & 779.872 - 686.667 i & 779.878 - 686.666 i  \\
\hline
 300      & 935.863 - 823.974 i   & 0 - 0.004165 i & 935.843 - 824.001 i & 935.848 - 824.000 i  \\ \hline
\end{tabular}
\end{center}
Table III. The fundamental quasinormal frequencies, $l=2$, $n=0$ for large
5-dimensional SAdS black hole.
\end{table}

\section{Charge}

When perturbing a charged black hole there excited two kinds of
scalar-type and vector-type gravitational modes: first induced
by the potential $V_{+}$ reducing at
$Q=0$ to the potential for electromagnetic perturbations, and second,
induced by $V_{-}$ reducing to the potential for purely gravitational
perturbations at $Q=0$. At $Q\neq 0$, which is the case of RN BH
background, there is no quasinormal mode which is purely electromagnetic
or purely gravitational; each quasinormal mode will be accompanied by
the emission of both electromagnetic and gravitational radiation.
Another interesting property of the RN BH ringing was observed in
\cite{onozawa}: it appeared that the quasinormal modes of
gravitational waves with multi-pole index $l$ coincide with
those of electromagnetic waves with multi-pole index $l-1$ in the extremal limit.
Then it was observed that this is connected with the fact that
the extremally charged RN black hole preserves super-symmetry, and
thus responses in a similar way on fields of different spin \cite{onozawa2}.
We do not observe similar effect in the case for the higher dimensional black
holes (see for example Table VII.)

In the charged background with $D>4$ we have generally two effective potentials
for scalar-type modes, two potentials for vector modes, and
one potential for tensor-type modes \cite{Ishibashi3}.
Yet the scalar-type potentials for some value of $l$ and $D$ have,
similar to the case of the neutral black hole, negative pitch near
the black hole horizon.

In order to compare the results obtained through the 6th order WKB
approximation and numerical QN modes for charged black hole we
give in the Appendix the fundamental QN modes for 4-dimensional RN
black hole. The agreement for tensor and $V_{+}$ vector modes with
numerical data is very good, while for $V_{-}$ the agreement is
worse. It is general for WKB approach that the more multi-pole index
$l$, the better accuracy it provides.

\begin{minipage}[c]{15cm}
\begin{longtable}{|r|r|r|r|}
\hline
  Q/M     & ~~tensor-type~~~~ & vector-type $V_{-}$~~~~&vector-type  $V_{+}$~~~~\\ \hline
  0.2   & 1.07084 - 0.25259 i & 0.80166 - 0.23717 i & 1.04554 - 0.24930 i \\
  0.4   & 1.07942 - 0.25139 i & 0.80271 - 0.23585 i  & 1.06791 - 0.24927  i \\
  0.5   & 1.08623 - 0.25028 i & 0.80363 - 0.23471 i & 1.08579 - 0.24898 i \\
  0.7   & 1.10613 - 0.24608 i & 0.80643 - 0.23094 i & 1.13963 - 0.24668 i \\
  0.8   & 1.12007 - 0.24209 i & 0.80816 - 0.22799 i & 1.17960 - 0.24329 i \\
  0.99  & 1.15571 - 0.22565 i & 0.81072 - 0.22002 i & 1.30465 - 0.21604 i \\
  $ext$ & 1.15786 - 0.22424 i & 0.81082 - 0.21953 i & 1.31425 - 0.21213 i \\ \hline
\end{longtable}
\centerline{}
Table IV. The fundamental ($n=0$) quasinormal frequencies, $l=2$,
for vector and tensor perturbations of 5-dimensional RN black hole.
$^{*}ext$ is near extremal value of $Q$ at which the QN mode converge to some
its limiting value (usually it is $Q=0.999999 M$ or closer to 1 M).
\end{minipage}

\begin{minipage}[c]{15cm}
\begin{longtable}{|r|r|r|r|}
\hline
  Q/M & ~~tensor-type~~~~ & vector-type $V_{-}$~~~~&vector-type  $V_{+}$~~~~\\ \hline
  0.2 & 1.59863 - 0.39798 i & 1.2255 - 0.381188 i & 1.57851 - 0.39361 i \\
  0.4 & 1.60531 - 0.39551 i & 1.22485 - 0.378598 i  & 1.60416 - 0.39225 i \\
  0.5 & 1.61054 - 0.39333 i & 1.22429 - 0.376596 i & 1.62477 - 0.39075 i \\
  0.7 & 1.62528 - 0.38578 i & 1.22237 - 0.371098 i & 1.68592 - 0.38487 i \\
  0.8 & 1.63479 - 0.37950 i & 1.22077 - 0.367652 i & 1.72882 - 0.37972 i \\
  0.99 & 1.65469 - 0.36053 i  & 1.21656 - 0.360428 i & 1.85359 - 0.34648 i \\
  $ext$ & 1.65572 - 0.35926 i & 1.21633 - 0.360039 i & 1.86325 - 0.34165 i \\ \hline
\end{longtable}
\centerline{}
Table V. The fundamental ($n=0$) quasinormal frequencies, $l=2$,
for vector and tensor perturbations of 6-dimensional RN black hole.
$^{*}ext$ is near extremal value of $Q$ at which the QN mode converge to some
its limiting value.
\end{minipage}

\begin{minipage}[c]{15cm}
\begin{longtable}{|r|r|r|r|}
\hline
  Q/M & ~~tensor-type~~~~ & vector-type $V_{-}$~~~~&vector-type  $V_{+}$~~~~\\ \hline
  0.2 & 2.10134 - 0.53092 i & 1.65365 - 0.509963 i & 2.08732 - 0.52520 i \\
  0.4 & 2.10623 - 0.52798 i & 1.65193 - 0.506922 i  & 2.11625 - 0.52210 i \\
  0.5 & 2.11012 - 0.52525 i & 1.65054 - 0.504657 i & 2.14023 - 0.51872 i \\
  0.7 & 2.12073 - 0.51570 i & 1.64641 - 0.498804 i & 2.21054 - 0.50807 i \\
  0.8 & 2.12678 - 0.50825 i & 1.64359 - 0.495383 i & 2.24964 - 0.50566 i \\
  0.99 &  2.13650 - 0.48935 i & 1.63746 - 0.488578 i & 2.32953 - 0.50006 i \\
  $ext$ & 2.13692 - 0.48824 i & 1.63714 - 0.488215 i & 2.34040 - 0.494896 i \\ \hline
\end{longtable}
\centerline{}
Table VI. The fundamental ($n=0$) quasinormal frequencies, $l=2$,
for vector and tensor perturbations of 7-dimensional RN black hole.
$^{*}ext$ is near extremal value of $Q$ at which the QN mode converge to some
its limiting value.
\end{minipage}

\begin{minipage}[c]{15cm}
\begin{longtable}{|r|r|r|r|}
\hline
  l & ~~tensor-type~~~~ & vector-type $V_{-}$~~~~&vector-type  $V_{+}$~~~~\\ \hline
  3 & 2.10134 - 0.53092 i & 0.23093 - 0.21529 i & 1.70850 - 0.21517 i \\
  4 & 2.10623 - 0.52798 i & 1.63511 - 0.21715 i  & 2.10087 - 0.21727 i \\
  5 & 2.11012 - 0.52525 i & 2.03228 - 0.21851 i & 2.49163 - 0.21861 i \\ \hline
\end{longtable}
\centerline{}
Table VII. The fundamental ($n=0$) quasinormal frequencies,
for vector and tensor perturbations of 5-dimensional extremal RN black hole.
\end{minipage}

\section{Appendix. Quasinormal modes of $D=4$ RN BH}

\begin{minipage}[c]{15cm}
\begin{longtable}{|r|r|r|r|}
\hline
  Q/M & ~~~numerical~~~~~ & third order WKB & six order WKB \\ \hline
  0.2 & 0.46296 - 0.09537 i & 0.46251 - 0.09543 i & 0.46296 - 0.09538 i \\
  0.4 & 0.47993 - 0.09644 i &0.47949 - 0.09649 i  & 0.47992 - 0.09645 i \\
  0.5 & 0.49368 - 0.09719 i & 0.49325 - 0.09723 i & 0.49367 - 0.09720 i \\
  0.7 & 0.53651 - 0.09877 i & 0.53613 - 0.09879 i & 0.53651 - 0.09878 i \\
  0.8 & 0.57013 - 0.09907 & 0.56978 - 0.09908 i & 0.57013 - 0.09908 i \\
  0.99 & 0.69275 - 0.08864 i & 0.69243 - 0.08863 i & 0.69275 - 0.08864 i \\
  $ext$ &  $0.70430 - 0.085973 i^{*}$ & 0.70398 - 0.08596 i & 0.704305 - 0.085972 i \\ \hline
\end{longtable}
\centerline{}
Table III. The fundamental quasinormal frequencies, $l=2$, $n=0$, $Z_{+}$
4-dimensional RN black hole; $^{*}ext$ is near extremal value of $Q$ at which
the QN mode converge to some
its limiting value (usually it is $Q=0.999999 M$ or closer to 1 M).
The numerical results for this limiting value is taken from the paper of H.Onozawa
et. al. \cite{onozawa}.
\end{minipage}

\begin{minipage}[c]{14cm}
\begin{longtable}{|r|r|r|r|}
\hline
  Q/M & ~~~numerical~~~~~~ & third order WKB & six order WKB \\ \hline
  0.2 & 0.37475 - 0.08907 i & 0.37423 - 0.08933 i & 0.37469 - 0.08900 i \\
  0.4 & 0.37844 - 0.08940 i & 0.37792 - 0.08965 i  & 0.37838 - 0.08933 i \\
  0.5 & 0.38168 - 0.08961 i & 0.38115 - 0.08985 i & 0.38162 - 0.08954 i \\
  0.7 & 0.39250 - 0.08990 i & 0.39191 - 0.09009 i & 0.39248 - 0.08982 i \\
  0.8 & 0.40122 - 0.08964 i & 0.40053 - 0.08978 i & 0.40125 - 0.08953 i \\
  0.99 &  0.42930 - 0.08427 i  & 0.42810 - 0.08431 i & 0.42955 - 0.08396 i \\
  $ext$ & $0.43134 - 0.083460 i^{*}$ & 0.43013 - 0.08349 i & 0.431612 - 0.083139 i \\ \hline
\end{longtable}
\centerline{}
Table III. The fundamental quasinormal frequencies, $l=2$, $n=0$, $Z_{-}$
4-dimensional RN black hole; $^{*}ext$ is near extremal value of $Q$ at which
the QN mode converge to some
its limiting value.
The numerical results for this limiting value is taken from the paper of H.Onozawa
et. al. \cite{onozawa}.
\end{minipage}

\section{Discussion}

We have been studied the gravitational quasinormal frequencies
for higher dimensional S, SdS, and SAdS black holes. The
quasinormal behavior crucially different in these three cases:
for $\Lambda=0$ the quasinormal modes are inversely proportional
to the black hole size (the horizon radius), the presence of
$\Lambda > 0$ leads to decreasing of the real oscillation frequency
and to increasing of the damping time. The presence of
$\Lambda < 0$ lead to totally different behavior: the
quasinormal modes of large and intermediate AdS black holes are
proportional to the black hole size (and thereby proportional to
the temperature in this regime), and goes to pure AdS limit when
the radius of the event horizon goes to zero \cite{konoplya}.
In all these points the higher dimensional gravitational modes
mimic the behavior of the four dimensional modes.
Different features in quasinormal behavior of the
higher dimensional black holes is connected with existence of the
three types of perturbations: scalar, vector and tensor types of gravitational
modes. These three types of perturbations excite three different
spectra of quasinormal modes. In particular, the more the spin weight
of the corresponding harmonics in which the perturbations are
expanded, the greater the real oscillation frequencies and the
greater the damping rate. Thus that is
the scalar-type modes which damp most slowly and thereby must
dominate at later stages of quasinormal ringing. And this is the case
of charged background, and also asymptotically de-Sitter or Anti-de-Sitter
backgrounds. All found here modes are damping what supports the
stability of the above black holes.

Unfortunately, with the help of existent semi-analytical methods,
we are not able to find the quasinormal modes of
scalar-type gravitational perturbations for a full range of
parameters, since the corresponding effective potentials have
negative pitch near the black hole horizon. Nevertheless in some
of these cases (such as the case of large $D$) one can anticipate
the quasinormal behavior.

Another interesting step in the direction 
of investingating of gravitational radiation
from higher dimensional black holes is the considering the
quantum effects such as hawking radiation (see \cite{kanti1} and references 
therein).

\section{Acknowledgments}
I wish to thank Akihiro Ishibashi for stimulating discussions.


\end{document}